\documentclass[twocolumn,superscriptaddress,showpacs,preprintnumbers,amsmath,amssymb,prl]{revtex4}
\usepackage{graphicx}
\usepackage{bm}
\usepackage[dvips]{color} 

\begin{document}


\title{Muon-spin-spectroscopy study of the penetration depth of FeTe$_{0.5}$Se$_{0.5}$}

\author{P.K. Biswas}
\affiliation{Physics Department, University of Warwick, Coventry, CV4 7AL, United Kingdom}

\author{G. Balakrishnan}
\affiliation{Physics Department, University of Warwick, Coventry, CV4 7AL, United Kingdom}

\author{D.M. Paul}
\affiliation{Physics Department, University of Warwick, Coventry, CV4 7AL, United Kingdom}

\author{C.V. Tomy}
\altaffiliation{Department of Physics, Indian Institute of Technology Bombay, Mumbai 40076, India}
\affiliation{Physics Department, University of Warwick, Coventry, CV4 7AL, United Kingdom}

\author{M.R. Lees}
\email[]{m.r.lees@warwick.ac.uk}
\affiliation{Physics Department, University of Warwick, Coventry, CV4 7AL, United Kingdom}

\author{A.D. Hillier}
\affiliation{ISIS Facility, Rutherford Appleton Laboratory, Chilton, Oxfordshire, OX11 0QX, U.K.}

\date{\today}

\begin{abstract}
Muon-spin-spectroscopy measurements have been used to study the superconducting state of FeTe$_{0.5}$Se$_{0.5}$. The temperature dependence of the in-plane magnetic penetration depth, $\lambda_{ab}\left(T\right)$, is found to be compatible with either a two gap $s+s$ wave or an anisotropic $s$-wave model. The value for $\lambda_{ab}\left(T\right)$ at $T=0$~K is estimated to be $\lambda_{ab}\left(0\right)=534(2)$~nm.
\end{abstract}

\maketitle

The discovery of superconductivity at high temperature in iron-based materials was a real surprise and has generated tremendous interest~\cite{Kamihara,Takahashi,Chen}. Iron selenium has recently been reported to be superconducting with a $T_c$ of 8.0~K at ambient pressure~\cite{Hsu} and 37~K at 7~GPa~\cite{Margadonna}. The substitution of tellurium on the selenium site increases $T_c$ to maximum of 14.5~K at ambient pressure.~\cite{Yeh,Fang} Magnetization and resistivity measurements indicate a lower critical field at $T=0$~K, $\mu_0H_{c1}(0)$, of between 1 and 8~mT and an upper critical field, $\mu_0H_{c2}(0)$, of 40-60~T for the FeTe$_{1-x}$Se$_{x}~\left(0.25\leq x\leq0.5\right)$ system.~\cite{Kida,Yadav,Biswas} Measurements on single crystals indicate that the superconducting properties of this material are anisotropic.~\cite{Yadav,Biswas}

Muon spin rotation/relaxation, ($\mu$SR), a probe that is sensitive to the local field distribution within a material, has often been used to measure the value and temperature dependence of the London magnetic penetration depth, $\lambda$, in the vortex state of type-II superconductors~\cite{Sonier, Brandt}. $\lambda^2\left(T\right)$ is in turn proportional to $n_s\left(T\right)$ where $n_s$ is the density of superconducting carriers.  The temperature and field dependence of $n_s$ can provide information on the nature of the superconducting gap. 

Here we report a $\mu$SR study of FeTe$_{0.5}$Se$_{0.5}$. We show that the temperature dependence of $\lambda$ can be equally well described using either a two gap $s+s$-wave or an anisotropic $s$-wave model. We obtain an in-plane magnetic penetration depth  $\lambda_{ab}(0)=534(2)$ nm. We compare these results with published data for the FeTe$_{1-x}$Se$_{x}\left(0.25\leq x\leq1.0\right)$ system. 

A polycrystalline sample of FeTe$_{0.5}$Se$_{0.5}$ was synthesized in a two-step process from high purity iron granules ($99.999\%$), selenium shot ($99.997\%$) and tellurium powder ($99.999\%$). First, the appropriate stoichiometric mixture of elements were sealed in an evacuated carbon coated quartz tube ($10^{-6}$ mbar) and heated at a rate of $100~^o$C/h to $650~^o$C, held at this temperature for 48 h, and then cooled to room temperature at $50~^o$C/h. The sample was then heated at a rate of $180~^o$C/h to $970~^o$C for 24 h and cooled to room temperature at $3~^o$C/h. Since the quartz tube often cracked during this cooling, the tube was sealed into a second quartz tube at a high vacuum before the second heating process.

\begin{figure}[tb]
\begin{center}
\includegraphics[width=0.9\columnwidth]{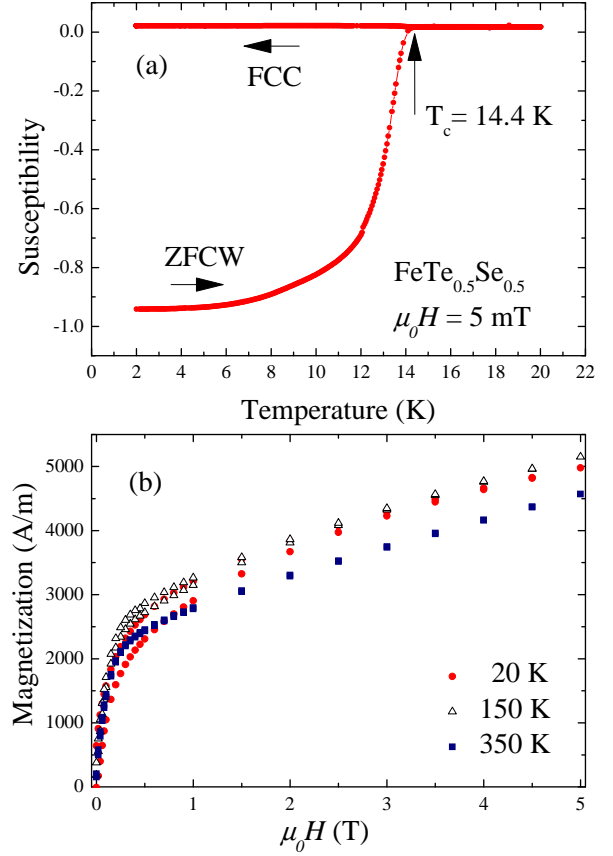}
\caption{\label{Figure1Biswas} (Color online) (a) Temperature dependence of the magnetic susceptibility of FeTe$_{0.5}$Se$_{0.5}$ measured using zero-field-cooled warming (ZFCW) and field-cooled cooling (FCC) protocols. The diamagnetic susceptibility corresponds to complete diamagnetic screening with a $T_c$ onset of 14.4~K. (b) Magnetization versus applied field curves for FeTe$_{0.5}$Se$_{0.5}$ collected above $T_c$ at 20~K, 150~K, and 350 K.}
\end{center}
\end{figure} 

X-ray diffraction measurements show the sample has a tetragonal PbO type structure (space group $P4/nmm$) with $a=b=0.3798(8)$~nm and $c=0.6042(12)$~nm. The x-ray pattern showed that the sample also contained a quantity ($<2\%$) of the hexagonal-NiAs phase of FeTe$_{0.5}$Se$_{0.5}$ and some Fe$_3$O$_4$ ($\sim1\%$). Magnetization ($M$) versus temperature ($T$) measurements carried out in a Quantum Design MPMS magnetometer reveal the sample has a transition temperature, $T_c$, of 14.4~K [see Fig.~\ref{Figure1Biswas}a]. The zero-field (ZF)-cooled dc susceptibility approaches a value of -1 while the field cooled signal is $10^{-3}$ indicating strong pinning in the sample. Magnetization versus applied magnetic field loops collected in the normal state at 20, 150 and 350~K [see Fig.~\ref{Figure1Biswas}b] show that the normal state signal is nearly temperature independent and has a response made up of contributions typical of a soft ferromagnet and a Pauli paramagnet. This is consistent with the presence in the sample of a small amount of Fe$_3$O$_4$. Using the published value for the saturation magnetization of Fe$_3$O$_4$~\cite{Walz} we estimate this fraction to be $\sim0.5\%$ of the sample by mass~\cite{KhasanovNote} in agreement with our x-ray diffraction measurements.

The $\mu$SR measurements were performed using the MuSR spectrometer at ISIS. A pulse of muons is produced every 20~ms and has a FWHM of $\sim70$~ns. The muons are implanted into the sample and decay with half life, $\tau_\mu$, of 2.2~$\mu$s into positrons, which are emitted preferentially in the direction of the muon spin axis. For a longitudinal-field (LF) measurement these positrons are detected and time stamped in the detectors which are positioned before ($F$) and after ($B$) sample. The positron counts $N_{F,B}\left(t\right)$ have the functional form
\begin{equation}
N_{F,B}\left(t\right)=N_{F,B}\left(0\right)\exp\left(\frac{-t}{\tau_\mu}\right)\left[1\pm G_Z\left(t\right)\right],
\end{equation}
where $G_Z\left(t\right)$ is the longitudinal relaxation function. $G_Z\left(t\right)$ is determined using 
\begin{equation}
\frac{N_{F}\left(t\right)-\alpha N_{B}\left(t\right)}{N_{F}\left(t\right)+\alpha N_{B}\left(t\right)},
\end{equation}
where $\alpha$ is a calibration constant which is determined by applying a transverse-field (TF) of 20 mT. Measurements were also carried out in transverse-field mode in magnetic fields of up to 60 mT. Each detector is normalized for the muon decay and rotated into two components at 90 degrees to one another. 

The powder sample (30 mm by 30 mm square and 1 mm thick) was mixed with GE varnish and mounted on a pure Ag plate. For measurements down to 1.2~K the sample was placed in a conventional Oxford Instruments cryostat. Data were also collected between 0.3 and 1~K in an Oxford Instruments He-3 cryostat. For the measurements in TF mode in the conventional cryostat, hematite slabs were positioned immediately behind the sample. For measurements in the He-3 cryostat these hematite slabs were removed to ensure good thermal contact between the sample and the cold stage of the cryostat leading to an increased background in the data collected. For all the data collected in a magnetic field presented in this paper, the sample was field-cooled to base temperature and the data collected while warming the sample in a field. A set of data collected at 1.2~K in an applied magnetic field, $\mu_{0}H=40$~mT after the sample was zero-field cooled produced no usable signal due to the very strong pinning present in the sample.  

\begin{figure}[tb]
\begin{center}
\includegraphics[width=0.9\columnwidth]{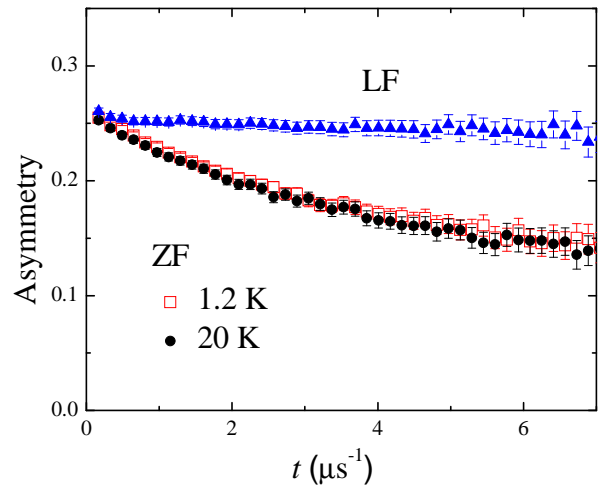}
\caption{\label{Figure2Biswas} (Color online) Zero-field ($T$=20 and 1.2~K) and longitudinal-field ($T$=20~K and $\mu_{0}H=30$~mT) $\mu$SR time spectra for a sample of FeTe$_{0.5}$Se$_{0.5}$.}
\end{center}
\end{figure}

Zero-field (ZF) $\mu$SR data (see Fig.~\ref{Figure2Biswas}) can be fitted using 

\begin{equation}
\label{Depolarization_Fit}
G_Z(t)=A_{0}\exp\left(-\Lambda t\right),
\end{equation}

where $A_0$ is the initial asymmetry, with a small nearly $T$ independent relaxation rate, $\Lambda$, of 0.19$\mu s^{-1}$ between 20~K and 1.2~K. The application of a small longitudinal magnetic field is sufficient to decouple the muon spin from the internal magnetic field. In line with the observations of Khasanov et al.~\cite{Khasanov} this suggests that the depolarization is caused by weak, static magnetic fields, that are present  in the sample both above and below $T_c$. The most likely source of this field is dilute, randomly oriented magnetic moments associated with the Fe$_3$O$_4$ impurity phase.

\begin{figure}[tb]
\begin{center}
\includegraphics[width=0.9\columnwidth]{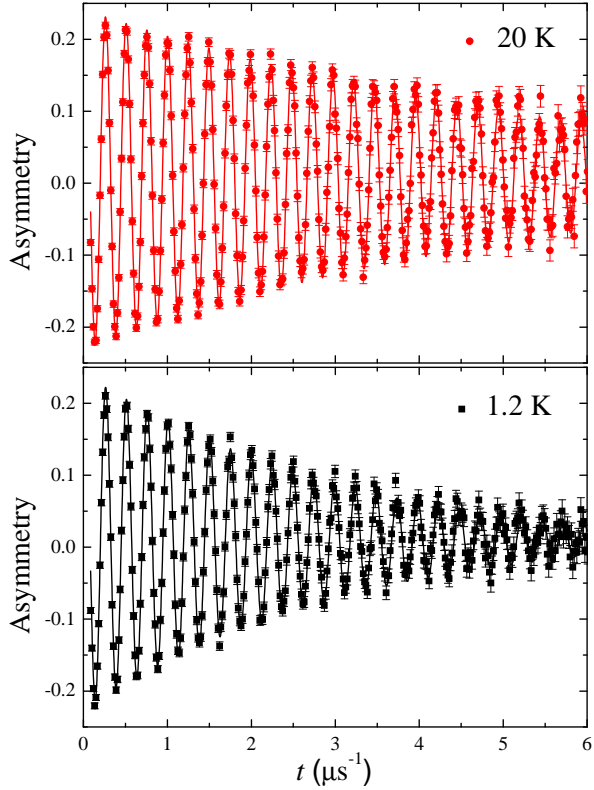}
\caption{\label{Figure3Biswas} (Color online) One component of the transverse-field muon-time spectra for FeTe$_{0.5}$Se$_{0.5}$ collected in a magnetic field $\mu_{0}H=30$~mT at temperatures above ($T=20$~K) and below ($T=1.2$~K) the superconducting transition temperature $T_c=14.4$~K.}
\end{center}
\end{figure}

\begin{figure}[tb]
\begin{center}
\includegraphics[width=1.0\columnwidth]{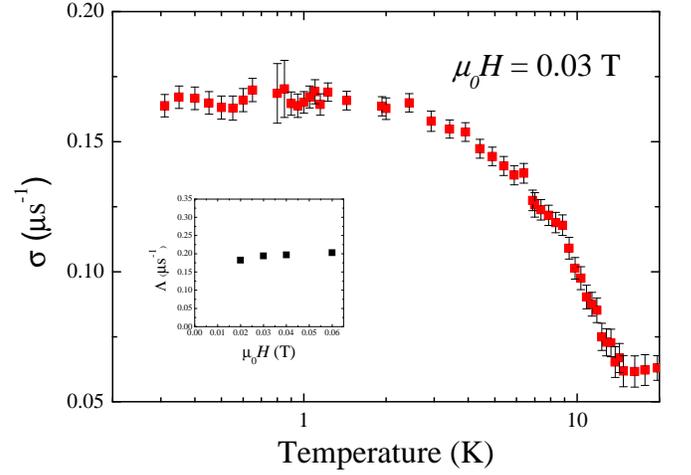}
\caption{\label{Figure4Biswas} (Color online)  The temperature dependence of the Gaussian depolarization rate $\sigma$ extracted from the TF muon-time spectra collected in an applied magnetic field $\mu_{0}H=30$~mT. The inset demonstrates the magnetic field independence of the parameter $\Lambda$ at 20~K.}
\end{center}
\end{figure}

Figure~\ref{Figure3Biswas} shows the TF-$\mu$SR precession signals above and below $T_c$. In the normal state, the oscillation shows a small relaxation. Below $T_c$, the relaxation rate increases due to the inhomogeneous field distribution of the flux line lattice. Previous measurements on polycrystalline samples of superconducting materials have shown that the internal field distributions can be modeled using a sinusoidally oscillating function with a Gaussian component 

\begin{equation}
\label{Depolarization_Fit}
G_X(t)=A_{0}\exp\left(-\Lambda t\right)\exp\left(-\sigma^{2}t^2\right)\cos\left(\omega t +\varphi\right),
\end{equation}

where $\omega$ is the muon precession frequency and $\varphi$ is the phase offset. $\sigma$ is the Gaussian relaxation rate given by  $\sigma=\left(\sigma^{2}_{sc} + \sigma^{2}_{nm}\right)^{\frac{1}{2}}$. $\sigma_{sc}\left(T\right)$ is the contribution to the relaxation arising from the vortex lattice while $\sigma_{nm}$, the nuclear magnetic dipolar term, is assumed to be temperature independent over the temperature range of the measurements. The data were fitted in two steps. First the data in the two channels were fitted simultaneously at each temperature with $A_0$, $\Lambda$, and $\sigma$ as common variables. The fits were checked over the entire temperature range to ensure that physical values were obtained for all the parameters at each temperature point. To ensure stability of the fits $\Lambda$ was then fixed to the value obtained just above $T_c$ and the data were refitted at each temperature point. The temperature dependence of $\sigma$ obtained is shown in Fig.~\ref{Figure4Biswas}.

In a superconductor with a large upper critical field and a hexagonal Abrikosov vortex lattice, the Gaussian muon-spin depolarization rate, $\sigma_{sc}\left(T\right)$, is related to the penetration depth $\lambda$ by the expression

\begin{equation}
\frac{2\sigma^2_{sc}\left(T\right)}{\gamma^{2}_{\mu}}=0.00371\frac{\Phi^{2}_{0}}{\lambda^4\left(T\right)}
\end{equation}

where $\gamma_{\mu}/2\pi=135.5$~MHz/T is the muon gyromagnetic ratio and $\Phi_{0}= 2.068 \times10^{-15}$~Wb is the flux quantum~\cite{Sonier, Brandt}. The temperature dependence of the penetration depth can then be fitted using either a single gap or a two gap model~\cite{Carrington, Padamsee}

\begin{equation}
\label{two_gap}
\frac{\lambda^{-2}\left(T\right)}{\lambda^{-2}\left(0\right)}=\omega_1\frac{\lambda^{-2}\left(T, \Delta_{0,1}\right)}{\lambda^{-2}\left(0,\Delta_{0,1}\right)}+\omega_2\frac{\lambda^{-2}\left(T, \Delta_{0,2}\right)}{\lambda^{-2}\left(0,\Delta_{0,2}\right)},
\end{equation}

where $\lambda^{-2}\left(0\right)$ is the value of the penetration depth at $T=0$~K, $\Delta_{0,i}$ is the value of the $i$-th ($i=1$ or 2) superconducting gap at $T=0$~K and $\omega_i$ is the weighting factor with $\omega_1+\omega_2=1$.

Each term in equation~\ref{two_gap} is evaluated using the standard expression within the local London approximation~\cite{Tinkham,Prozorov}

\begin{equation}
\frac{\lambda^{-2}\left(T, \Delta_{0,i}\right)}{\lambda^{-2}\left(0, \Delta_{0,i}\right)}=1+\frac{1}{\pi}\int^{2\pi}_{0}\int^{\infty}_{\Delta_{\left(T,\varphi\right)}}\left(\frac{\partial f}{\partial E}\right)\frac{ EdE d\varphi}{\sqrt{E^2-\Delta_i\left(T,\varphi\right)^2}},
\end{equation}

where $f=\left[1+\exp\left(E/k_BT\right)\right]^{-1}$ is the Fermi function, $\varphi$ is the angle along the Fermi surface, and $\Delta_i\left(T,\varphi\right)=\Delta_{0, i}\delta\left(T/T_c\right)g\left(\varphi\right)$. The temperature dependence of the gap is approximated by the expression $\delta\left(T/T_c\right)=\tanh\left\{1.82\left[1.018\left(T_c/T-1\right)\right]^{0.51}\right\}$~\cite{Carrington} while $g\left(\varphi\right)$ describes the angular dependence of the gap. The fits (see Table~$\ref{table_of_fits}$) appear to rule out both the $d$-wave and $s$-wave as possible models for this system~\cite{Errors1}.

\begin{table*}
\caption{Results for fits to the temperature dependence of the penetration depth using different models for the symmetry of the superconducting gap function~\cite{Errors1}.}
\label{table_of_fits}
\begin{center}
\begin{tabular}[t]{|l|l|l|c|}\hline
Model&$g\left(\varphi\right)$& Gap value (meV)&$\chi ^2$\\\hline
$s$-wave&1&$\Delta$=1.86(2)&5.93\\\hline
$s+s$-wave&1&$\Delta_1$=2.6(1), $\Delta_2$=0.87(6) and $\omega_1=0.70(3)$&1.55\\\hline
anisotropic $s$-wave&$\left(s+\cos4\varphi\right)$&$\Delta=1.4(1)$ with $s=1.56(5)$&1.62\\\hline
$d$-wave&$\left|\cos\left(2\varphi\right)\right|$&$\Delta=3.31(4)$&2.87\\\hline
\end{tabular}
\par\medskip\footnotesize
\end{center}
\end{table*} 

The anisotropic $s$-wave model gives a value for $s$, the parameter reflecting the isotropic $s$-wave component, that is larger than that obtained for FeSe$_{0.85}$ in Ref.~\cite{Khasanov}. Nevertheless, the variation in the gap with angle $\Delta_{max}/\Delta_{min}\approx4.6$ is still larger than the published values for related single layer superconductor NdFeAsO$_{0.9}$F$_{0.1}$.~\cite{Kondo} 

\begin{figure}[tb]
\begin{center}
\includegraphics[width=0.9\columnwidth]{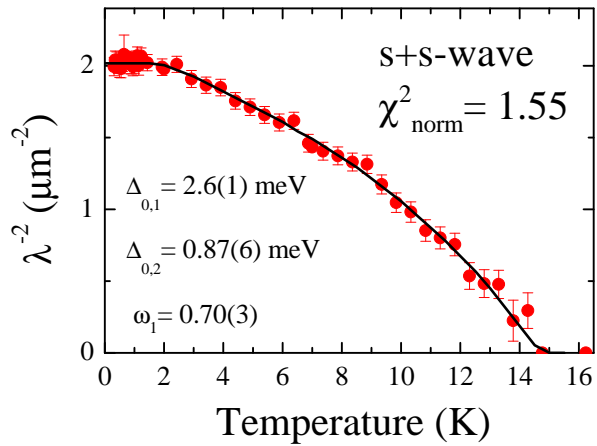}
\caption{\label{Figure5Biswas} (Color online) Temperature dependence of $\lambda^{-2}$ for FeTe$_{0.5}$Se$_{0.5}$. The curve (black line) is a fit to the data using two $s$-wave components, each with an isotropic gap.}
\end{center}
\end{figure}

A fit to the data using a two-gap $s+s$ wave model is shown in Fig.~\ref{Figure5Biswas} and gives $\Delta_{0,1}=2.6(1)$~meV and $\Delta_{0,2}=0.87(6)$~meV with $\omega_1=0.70(3)$.  This model gives the lowest $\chi^2$. $\omega_1$ agrees with the value obtained by $\mu$SR for FeSe$_{0.85}$ where $\omega_1=0.658(3)$.~\cite{Khasanov} The size of the larger energy gap for FeSe$_{0.85}$ and FeTe$_{0.5}$Se$_{0.5}$ scale with $T_c$. The ratio $\Delta_{0,1}/\Delta_{0,2}\sim3$ found in FeTe$_{0.5}$Se$_{0.5}$ is $40\%$ smaller than the corresponding value seen in FeSe$_{0.85}$ but coincides with the value for $R$FeAsO$_{0.9}$F$_{0.1}$ ($R$=La, Nd) determined by measuring the magnetic penetration depth using a tunnel-diode resonator.~\cite{Martin}. 

For anisotropic polycrystalline samples, the magnetic penetration depth, $\lambda$, calculated from the $\mu$SR depolarization rate $\sigma$ is related to  $\lambda_{ab}$, the in-plane penetration depth by $\lambda=3^{\frac{1}{4}}\lambda_{ab}$.~\cite{LambdaNote, Fesenko} At $T=0$, the value for $\lambda\left(0\right)=703(2)$~nm with $\lambda_{ab}\left(0\right)=534(2)$~nm~\cite{Errors2}. These values are longer than those obtained by Khasanov $\textit{et al}$.~\cite{Khasanov} for FeSe$_{0.85}$ in spite of the fact that the $T_c$ of FeTe$_{0.5}$Se$_{0.5}$ is $\sim6$~K higher. The value for $\lambda$ would place FeTe$_{0.5}$Se$_{0.5}$ above the line for hole doped high $T_c$ cuprates on an Uemura plot \cite{Uemura,Luetkens}.

Using an upper critical field, $B_{c2}\left(0\right)\parallel ab$, for FeTe$_{0.5}$Se$_{0.5}$ estimated from transport and magnetization measurements of 50~T and $B_{c2}=\frac{\Phi_{0}}{2\pi\xi^2}$, we calculate a coherence length, $\xi_{ab}$, for FeTe$_{0.5}$Se$_{0.5}$ at 0 K of $\sim2.6$~nm. If this is combined with our measurement of $\lambda$ and the standard expression $H_{c1}=\frac{\Phi_{0}}{4\pi\lambda^2}\left(\ln\frac{\lambda}{\xi}+0.12\right)$~\cite{Tinkham} we estimate $\mu_{0}H_{c1}\left(0\right)\parallel ab=3.2$~mT. This is in fair agreement with magnetization measurements where the first deviation from linear behavior gives $\mu_{0}H_{c1}//ab$ of between 1 and 8 mT at 1.5~K.~\cite{Yadav, Biswas} 

In summary, $\mu$SR measurements have been performed on superconducting FeTe$_{0.5}$Se$_{0.5}$.  The temperature dependence of the magnetic penetration depth is found to be compatible with either a two-gap $s+s$-wave or an anisotropic $s$-wave model. This result is consistent with other experimental data~\cite{Khasanov, Dong} and density-functional calculations of the physical properties~\cite{Subedi} of the FeTe$_{1-x}$Se$_{x}$ system, and with the more general picture that is now emerging for iron pnictide superconductors in which they are described as two-band superconductors. Further studies on high purity single crystal samples are desirable as the presence of impurities can sometimes mask the true nature of the superconducting gap.~\cite {Sonier2}

\begin{acknowledgments}
The Quantum Design MPMS magnetometer used in this research was obtained through the Science City Advanced Materials project: Creating and Characterising Next Generation Advanced Materials project, with support from Advantage West Midlands (AWM) and part funded by the European Regional Development Fund (ERDF).     
\end{acknowledgments}

\bibliography{paper091509}

\begin{thebibliography}{30}
\expandafter\ifx\csname natexlab\endcsname\relax\def\natexlab#1{#1}\fi
\expandafter\ifx\csname bibnamefont\endcsname\relax
  \def\bibnamefont#1{#1}\fi
\expandafter\ifx\csname bibfnamefont\endcsname\relax
  \def\bibfnamefont#1{#1}\fi
\expandafter\ifx\csname citenamefont\endcsname\relax
  \def\citenamefont#1{#1}\fi
\expandafter\ifx\csname url\endcsname\relax
  \def\url#1{\texttt{#1}}\fi
\expandafter\ifx\csname urlprefix\endcsname\relax\def\urlprefix{URL }\fi
\providecommand{\bibinfo}[2]{#2}
\providecommand{\eprint}[2][]{\url{#2}}

\bibitem[{\citenamefont{Kamihara et~al.}(2008)\citenamefont{Kamihara, Watanabe,
  Hirano, and Hosono}}]{Kamihara}
\bibinfo{author}{\bibfnamefont{Y.}~\bibnamefont{Kamihara}},
  \bibinfo{author}{\bibfnamefont{T.}~\bibnamefont{Watanabe}},
  \bibinfo{author}{\bibfnamefont{M.}~\bibnamefont{Hirano}}, \bibnamefont{and}
  \bibinfo{author}{\bibfnamefont{H.}~\bibnamefont{Hosono}},
  \bibinfo{journal}{J. Am. Chem. Soc.} \textbf{\bibinfo{volume}{130}},
  \bibinfo{pages}{3296} (\bibinfo{year}{2008}).

\bibitem[{\citenamefont{Takahashi et~al.}(2008)\citenamefont{Takahashi, Igawa,
  Arii, Kamihara, Hirano, and Hosono}}]{Takahashi}
\bibinfo{author}{\bibfnamefont{H.}~\bibnamefont{Takahashi}},
  \bibinfo{author}{\bibfnamefont{K.}~\bibnamefont{Igawa}},
  \bibinfo{author}{\bibfnamefont{K.}~\bibnamefont{Arii}},
  \bibinfo{author}{\bibfnamefont{Y.}~\bibnamefont{Kamihara}},
  \bibinfo{author}{\bibfnamefont{M.}~\bibnamefont{Hirano}}, \bibnamefont{and}
  \bibinfo{author}{\bibfnamefont{H.}~\bibnamefont{Hosono}},
  \bibinfo{journal}{Nature (London)} \textbf{\bibinfo{volume}{453}},
  \bibinfo{pages}{376} (\bibinfo{year}{2008}).

\bibitem[{\citenamefont{Chen et~al.}(2008)\citenamefont{Chen, Wu, Wu, Liu,
  Chen, and Fang}}]{Chen}
\bibinfo{author}{\bibfnamefont{X.~H.} \bibnamefont{Chen}},
  \bibinfo{author}{\bibfnamefont{T.}~\bibnamefont{Wu}},
  \bibinfo{author}{\bibfnamefont{G.}~\bibnamefont{Wu}},
  \bibinfo{author}{\bibfnamefont{R.~H.} \bibnamefont{Liu}},
  \bibinfo{author}{\bibfnamefont{H.}~\bibnamefont{Chen}}, \bibnamefont{and}
  \bibinfo{author}{\bibfnamefont{D.~F.} \bibnamefont{Fang}},
  \bibinfo{journal}{Nature (London)} \textbf{\bibinfo{volume}{453}},
  \bibinfo{pages}{761} (\bibinfo{year}{2008}).

\bibitem[{\citenamefont{Hsu et~al.}(2008)\citenamefont{Hsu, Luo, Yeh, Chen,
  Huang, Wu, Lee, Huang, Chu, Yan et~al.}}]{Hsu}
\bibinfo{author}{\bibfnamefont{F.-C.} \bibnamefont{Hsu}},
  \bibinfo{author}{\bibfnamefont{J.-Y.} \bibnamefont{Luo}},
  \bibinfo{author}{\bibfnamefont{K.-W.} \bibnamefont{Yeh}},
  \bibinfo{author}{\bibfnamefont{T.-K.} \bibnamefont{Chen}},
  \bibinfo{author}{\bibfnamefont{T.-W.} \bibnamefont{Huang}},
  \bibinfo{author}{\bibfnamefont{P.~M.} \bibnamefont{Wu}},
  \bibinfo{author}{\bibfnamefont{Y.-C.} \bibnamefont{Lee}},
  \bibinfo{author}{\bibfnamefont{Y.-L.} \bibnamefont{Huang}},
  \bibinfo{author}{\bibfnamefont{Y.-Y.} \bibnamefont{Chu}},
  \bibinfo{author}{\bibfnamefont{D.-C.} \bibnamefont{Yan}},
  \bibnamefont{et~al.}, \bibinfo{journal}{Proc. Natl. Acad. Sci.}
  \textbf{\bibinfo{volume}{105}}, \bibinfo{pages}{14262}
  (\bibinfo{year}{2008}).

\bibitem[{\citenamefont{Margadonna et~al.}(2009)\citenamefont{Margadonna,
  Takabayashi, Ohishi, Mizuguchi, Takano, Kagayama, Nakagawa, Takata, and
  Prassides}}]{Margadonna}
\bibinfo{author}{\bibfnamefont{S.}~\bibnamefont{Margadonna}},
  \bibinfo{author}{\bibfnamefont{Y.}~\bibnamefont{Takabayashi}},
  \bibinfo{author}{\bibfnamefont{Y.}~\bibnamefont{Ohishi}},
  \bibinfo{author}{\bibfnamefont{Y.}~\bibnamefont{Mizuguchi}},
  \bibinfo{author}{\bibfnamefont{Y.}~\bibnamefont{Takano}},
  \bibinfo{author}{\bibfnamefont{T.}~\bibnamefont{Kagayama}},
  \bibinfo{author}{\bibfnamefont{T.}~\bibnamefont{Nakagawa}},
  \bibinfo{author}{\bibfnamefont{M.}~\bibnamefont{Takata}}, \bibnamefont{and}
  \bibinfo{author}{\bibfnamefont{K.}~\bibnamefont{Prassides}},
  \bibinfo{journal}{Phys. Rev. B} \textbf{\bibinfo{volume}{80}},
  \bibinfo{pages}{064506} (\bibinfo{year}{2009}).

\bibitem[{\citenamefont{Yeh et~al.}(2008)\citenamefont{Yeh, Huang, Huang, Chen,
  Hsu, Wu, Lee, Chu, Chen, Luo et~al.}}]{Yeh}
\bibinfo{author}{\bibfnamefont{K.-W.} \bibnamefont{Yeh}},
  \bibinfo{author}{\bibfnamefont{T.-W.} \bibnamefont{Huang}},
  \bibinfo{author}{\bibfnamefont{Y.-l.} \bibnamefont{Huang}},
  \bibinfo{author}{\bibfnamefont{T.-K.} \bibnamefont{Chen}},
  \bibinfo{author}{\bibfnamefont{F.-C.} \bibnamefont{Hsu}},
  \bibinfo{author}{\bibfnamefont{P.~M.} \bibnamefont{Wu}},
  \bibinfo{author}{\bibfnamefont{Y.-C.} \bibnamefont{Lee}},
  \bibinfo{author}{\bibfnamefont{Y.-Y.} \bibnamefont{Chu}},
  \bibinfo{author}{\bibfnamefont{C.-L.} \bibnamefont{Chen}},
  \bibinfo{author}{\bibfnamefont{J.-Y.} \bibnamefont{Luo}},
  \bibnamefont{et~al.}, \bibinfo{journal}{Europhys. Lett.}
  \textbf{\bibinfo{volume}{84}}, \bibinfo{pages}{{37002}}
  (\bibinfo{year}{2008}).

\bibitem[{\citenamefont{Fang et~al.}(2008)\citenamefont{Fang, Pham, Qian, Liu,
  Vehstedt, Liu, Spinu, and Mao}}]{Fang}
\bibinfo{author}{\bibfnamefont{M.~H.} \bibnamefont{Fang}},
  \bibinfo{author}{\bibfnamefont{H.~M.} \bibnamefont{Pham}},
  \bibinfo{author}{\bibfnamefont{B.}~\bibnamefont{Qian}},
  \bibinfo{author}{\bibfnamefont{T.~J.} \bibnamefont{Liu}},
  \bibinfo{author}{\bibfnamefont{E.~K.} \bibnamefont{Vehstedt}},
  \bibinfo{author}{\bibfnamefont{Y.}~\bibnamefont{Liu}},
  \bibinfo{author}{\bibfnamefont{L.}~\bibnamefont{Spinu}}, \bibnamefont{and}
  \bibinfo{author}{\bibfnamefont{Z.~Q.} \bibnamefont{Mao}},
  \bibinfo{journal}{Phys. Rev. B} \textbf{\bibinfo{volume}{78}},
  \bibinfo{pages}{{224503}} (\bibinfo{year}{2008}).

\bibitem[{\citenamefont{Kida et~al.}(2009)\citenamefont{Kida, Matsunaga,
  Hagiwara, Mizuguchi, Takano, and Kindo}}]{Kida}
\bibinfo{author}{\bibfnamefont{T.}~\bibnamefont{Kida}},
  \bibinfo{author}{\bibfnamefont{T.}~\bibnamefont{Matsunaga}},
  \bibinfo{author}{\bibfnamefont{M.}~\bibnamefont{Hagiwara}},
  \bibinfo{author}{\bibfnamefont{Y.}~\bibnamefont{Mizuguchi}},
  \bibinfo{author}{\bibfnamefont{Y.}~\bibnamefont{Takano}}, \bibnamefont{and}
  \bibinfo{author}{\bibfnamefont{K.}~\bibnamefont{Kindo}}, \bibinfo{journal}{J.
  Phys. Soc. Japan} \textbf{\bibinfo{volume}{78}}, \bibinfo{pages}{113701}
  (\bibinfo{year}{2009}).

\bibitem[{\citenamefont{Yadav and Paulose}(2009)}]{Yadav}
\bibinfo{author}{\bibfnamefont{C.~S.} \bibnamefont{Yadav}} \bibnamefont{and}
  \bibinfo{author}{\bibfnamefont{P.~L.} \bibnamefont{Paulose}},
  \bibinfo{journal}{New J. Phys.} \textbf{\bibinfo{volume}{11}},
  \bibinfo{pages}{103046} (\bibinfo{year}{2009}).

\bibitem[{\citenamefont{Biswas et~al.}(2010)\citenamefont{Biswas, Balakrishnan,
  Tomy, Paul, and Lees}}]{Biswas}
\bibinfo{author}{\bibfnamefont{P.~K.} \bibnamefont{Biswas}},
  \bibinfo{author}{\bibfnamefont{G.}~\bibnamefont{Balakrishnan}},
  \bibinfo{author}{\bibfnamefont{C.~V.} \bibnamefont{Tomy}},
  \bibinfo{author}{\bibfnamefont{D.~M.} \bibnamefont{Paul}}, \bibnamefont{and}
  \bibinfo{author}{\bibfnamefont{M.~R.} \bibnamefont{Lees}}
  (\bibinfo{year}{2010}), \bibinfo{note}{unpublished}.

\bibitem[{\citenamefont{Sonier et~al.}(2000)\citenamefont{Sonier, Brewer, and
  Kiefl}}]{Sonier}
\bibinfo{author}{\bibfnamefont{J.~E.} \bibnamefont{Sonier}},
  \bibinfo{author}{\bibfnamefont{J.~H.} \bibnamefont{Brewer}},
  \bibnamefont{and} \bibinfo{author}{\bibfnamefont{R.~F.} \bibnamefont{Kiefl}},
  \bibinfo{journal}{Rev. Mod. Phys.} \textbf{\bibinfo{volume}{72}},
  \bibinfo{pages}{769} (\bibinfo{year}{2000}).

\bibitem[{\citenamefont{Brandt}(1988)}]{Brandt}
\bibinfo{author}{\bibfnamefont{E.~H.} \bibnamefont{Brandt}},
  \bibinfo{journal}{Phys. Rev. B} \textbf{\bibinfo{volume}{37}},
  \bibinfo{pages}{2349} (\bibinfo{year}{1988}).

\bibitem[{\citenamefont{Walz}(2002)}]{Walz}
\bibinfo{author}{\bibfnamefont{F.}~\bibnamefont{Walz}}, \bibinfo{journal}{J.
  Phys.: Condens. Matter} \textbf{\bibinfo{volume}{14}}, \bibinfo{pages}{R285}
  (\bibinfo{year}{2002}).

\bibitem[{Kha()}]{KhasanovNote}
\bibinfo{note}{The magnetization at 5~T and 20~K is $\sim30$ times smaller than
  that reported in the sample of FeSe$_{0.85}$ studied by Khasanov \textit{et
  al}.~\cite{Khasanov}}.

\bibitem[{\citenamefont{Khasanov et~al.}(2008)\citenamefont{Khasanov, Conder,
  Pomjakushina, Amato, Baines, Bukowski, Karpinski, Katrych, Klauss, Luetkens
  et~al.}}]{Khasanov}
\bibinfo{author}{\bibfnamefont{R.}~\bibnamefont{Khasanov}},
  \bibinfo{author}{\bibfnamefont{K.}~\bibnamefont{Conder}},
  \bibinfo{author}{\bibfnamefont{E.}~\bibnamefont{Pomjakushina}},
  \bibinfo{author}{\bibfnamefont{A.}~\bibnamefont{Amato}},
  \bibinfo{author}{\bibfnamefont{C.}~\bibnamefont{Baines}},
  \bibinfo{author}{\bibfnamefont{Z.}~\bibnamefont{Bukowski}},
  \bibinfo{author}{\bibfnamefont{J.}~\bibnamefont{Karpinski}},
  \bibinfo{author}{\bibfnamefont{S.}~\bibnamefont{Katrych}},
  \bibinfo{author}{\bibfnamefont{H.-H.} \bibnamefont{Klauss}},
  \bibinfo{author}{\bibfnamefont{H.}~\bibnamefont{Luetkens}},
  \bibnamefont{et~al.}, \bibinfo{journal}{Phys. Rev. B}
  \textbf{\bibinfo{volume}{78}}, \bibinfo{pages}{220510(R)}
  (\bibinfo{year}{2008}).

\bibitem[{\citenamefont{Carrington and Manzano}(2003)}]{Carrington}
\bibinfo{author}{\bibfnamefont{A.}~\bibnamefont{Carrington}} \bibnamefont{and}
  \bibinfo{author}{\bibfnamefont{F.}~\bibnamefont{Manzano}},
  \bibinfo{journal}{Physica C} \textbf{\bibinfo{volume}{385}},
  \bibinfo{pages}{205 } (\bibinfo{year}{2003}).

\bibitem[{\citenamefont{Padamsee et~al.}(1973)\citenamefont{Padamsee, Neighbor,
  and Shiffman}}]{Padamsee}
\bibinfo{author}{\bibfnamefont{H.}~\bibnamefont{Padamsee}},
  \bibinfo{author}{\bibfnamefont{J.~E.} \bibnamefont{Neighbor}},
  \bibnamefont{and} \bibinfo{author}{\bibfnamefont{C.~A.}
  \bibnamefont{Shiffman}}, \bibinfo{journal}{J. Low Temp. Phys.}
  \textbf{\bibinfo{volume}{12}}, \bibinfo{pages}{387 } (\bibinfo{year}{1973}).

\bibitem[{\citenamefont{Tinkham}(1975)}]{Tinkham}
\bibinfo{author}{\bibfnamefont{M.}~\bibnamefont{Tinkham}},
  \emph{\bibinfo{title}{Introduction to Superconductivity}}
  (\bibinfo{publisher}{McGraw-Hill, New York}, \bibinfo{year}{1975}).

\bibitem[{\citenamefont{Prozorov and Giannetta}(2006)}]{Prozorov}
\bibinfo{author}{\bibfnamefont{R.}~\bibnamefont{Prozorov}} \bibnamefont{and}
  \bibinfo{author}{\bibfnamefont{R.~W.} \bibnamefont{Giannetta}},
  \bibinfo{journal}{Superconductor Science and Technology}
  \textbf{\bibinfo{volume}{19}}, \bibinfo{pages}{R41} (\bibinfo{year}{2006}).

\bibitem[{Err({\natexlab{a}})}]{Errors1}
\bibinfo{note}{The normalized $\chi^2$ values, resulting from our least squares
  fits to the temperature dependence of $\lambda^{-2}$ using different models
  for the gap, are used as criteria to determine which model best describes the
  data.}

\bibitem[{\citenamefont{Kondo et~al.}(2008)\citenamefont{Kondo, Santander-Syro,
  Copie, Liu, Tillman, Mun, Schmalian, Bud'ko, Tanatar, Canfield
  et~al.}}]{Kondo}
\bibinfo{author}{\bibfnamefont{T.}~\bibnamefont{Kondo}},
  \bibinfo{author}{\bibfnamefont{A.~F.} \bibnamefont{Santander-Syro}},
  \bibinfo{author}{\bibfnamefont{O.}~\bibnamefont{Copie}},
  \bibinfo{author}{\bibfnamefont{C.}~\bibnamefont{Liu}},
  \bibinfo{author}{\bibfnamefont{M.~E.} \bibnamefont{Tillman}},
  \bibinfo{author}{\bibfnamefont{E.~D.} \bibnamefont{Mun}},
  \bibinfo{author}{\bibfnamefont{J.}~\bibnamefont{Schmalian}},
  \bibinfo{author}{\bibfnamefont{S.~L.} \bibnamefont{Bud'ko}},
  \bibinfo{author}{\bibfnamefont{M.~A.} \bibnamefont{Tanatar}},
  \bibinfo{author}{\bibfnamefont{P.~C.} \bibnamefont{Canfield}},
  \bibnamefont{et~al.}, \bibinfo{journal}{Phys. Rev. Lett.}
  \textbf{\bibinfo{volume}{101}}, \bibinfo{pages}{147003}
  (\bibinfo{year}{2008}).

\bibitem[{\citenamefont{Martin et~al.}(2009)\citenamefont{Martin, Tillman, Kim,
  Tanatar, Kim, Kreyssig, Gordon, Vannette, Nandi, Kogan et~al.}}]{Martin}
\bibinfo{author}{\bibfnamefont{C.}~\bibnamefont{Martin}},
  \bibinfo{author}{\bibfnamefont{M.~E.} \bibnamefont{Tillman}},
  \bibinfo{author}{\bibfnamefont{H.}~\bibnamefont{Kim}},
  \bibinfo{author}{\bibfnamefont{M.~A.} \bibnamefont{Tanatar}},
  \bibinfo{author}{\bibfnamefont{S.~K.} \bibnamefont{Kim}},
  \bibinfo{author}{\bibfnamefont{A.}~\bibnamefont{Kreyssig}},
  \bibinfo{author}{\bibfnamefont{R.~T.} \bibnamefont{Gordon}},
  \bibinfo{author}{\bibfnamefont{M.~D.} \bibnamefont{Vannette}},
  \bibinfo{author}{\bibfnamefont{S.}~\bibnamefont{Nandi}},
  \bibinfo{author}{\bibfnamefont{V.~G.} \bibnamefont{Kogan}},
  \bibnamefont{et~al.}, \bibinfo{journal}{Phys. Rev. Lett.}
  \textbf{\bibinfo{volume}{102}}, \bibinfo{pages}{247002}
  (\bibinfo{year}{2009}).

\bibitem[{\citenamefont{Fesenko et~al.}(1991)\citenamefont{Fesenko, Gorbunov,
  and Smilga}}]{Fesenko}
\bibinfo{author}{\bibfnamefont{V.}~\bibnamefont{Fesenko}},
  \bibinfo{author}{\bibfnamefont{V.}~\bibnamefont{Gorbunov}}, \bibnamefont{and}
  \bibinfo{author}{\bibfnamefont{V.}~\bibnamefont{Smilga}},
  \bibinfo{journal}{Physica C} \textbf{\bibinfo{volume}{176}},
  \bibinfo{pages}{551 } (\bibinfo{year}{1991}).

\bibitem[{Lam()}]{LambdaNote}
\bibinfo{note}{$\mu$SR cannot measure $\lambda$ along a single crystallographic
  direction. The measurements are limited to mixed quantities, which here is
  $\lambda_{ab}=\left(\lambda_{a}\lambda_{b}\right)^{1/2}$.}

\bibitem[{Err({\natexlab{b}})}]{Errors2}
\bibinfo{note}{The error in $\lambda(0)$ is the statistical error arising from
  the fit to the $\lambda^{-2}(T)$ data using the model described in the text.
  The error quoted does not take into account any systematic errors (e.g.
  vortex lattice disorder) that may be present in the data.}

\bibitem[{\citenamefont{Uemura et~al.}(1989)\citenamefont{Uemura, Luke,
  Sternlieb, Brewer, Carolan, Hardy, Kadono, Kempton, Kiefl, Kreitzman
  et~al.}}]{Uemura}
\bibinfo{author}{\bibfnamefont{Y.~J.} \bibnamefont{Uemura}},
  \bibinfo{author}{\bibfnamefont{G.~M.} \bibnamefont{Luke}},
  \bibinfo{author}{\bibfnamefont{B.~J.} \bibnamefont{Sternlieb}},
  \bibinfo{author}{\bibfnamefont{J.~H.} \bibnamefont{Brewer}},
  \bibinfo{author}{\bibfnamefont{J.~F.} \bibnamefont{Carolan}},
  \bibinfo{author}{\bibfnamefont{W.~N.} \bibnamefont{Hardy}},
  \bibinfo{author}{\bibfnamefont{R.}~\bibnamefont{Kadono}},
  \bibinfo{author}{\bibfnamefont{J.~R.} \bibnamefont{Kempton}},
  \bibinfo{author}{\bibfnamefont{R.~F.} \bibnamefont{Kiefl}},
  \bibinfo{author}{\bibfnamefont{S.~R.} \bibnamefont{Kreitzman}},
  \bibnamefont{et~al.}, \bibinfo{journal}{Phys. Rev. Lett.}
  \textbf{\bibinfo{volume}{62}}, \bibinfo{pages}{2317} (\bibinfo{year}{1989}).

\bibitem[{\citenamefont{Luetkens et~al.}(2008)\citenamefont{Luetkens, Klauss,
  Khasanov, Amato, Klingeler, Hellmann, Leps, Kondrat, Hess, K\"{o}hler
  et~al.}}]{Luetkens}
\bibinfo{author}{\bibfnamefont{H.}~\bibnamefont{Luetkens}},
  \bibinfo{author}{\bibfnamefont{H.-H.} \bibnamefont{Klauss}},
  \bibinfo{author}{\bibfnamefont{R.}~\bibnamefont{Khasanov}},
  \bibinfo{author}{\bibfnamefont{A.}~\bibnamefont{Amato}},
  \bibinfo{author}{\bibfnamefont{R.}~\bibnamefont{Klingeler}},
  \bibinfo{author}{\bibfnamefont{I.}~\bibnamefont{Hellmann}},
  \bibinfo{author}{\bibfnamefont{N.}~\bibnamefont{Leps}},
  \bibinfo{author}{\bibfnamefont{A.}~\bibnamefont{Kondrat}},
  \bibinfo{author}{\bibfnamefont{C.}~\bibnamefont{Hess}},
  \bibinfo{author}{\bibfnamefont{A.}~\bibnamefont{K\"{o}hler}},
  \bibnamefont{et~al.}, \bibinfo{journal}{Phys. Rev. Lett.}
  \textbf{\bibinfo{volume}{101}}, \bibinfo{pages}{097009}
  (\bibinfo{year}{2008}).

\bibitem[{\citenamefont{Dong et~al.}(2009)\citenamefont{Dong, Guan, Zhou, Qiu,
  Ding, Zhang, Patel, Xiao, and Li}}]{Dong}
\bibinfo{author}{\bibfnamefont{J.~K.} \bibnamefont{Dong}},
  \bibinfo{author}{\bibfnamefont{T.~Y.} \bibnamefont{Guan}},
  \bibinfo{author}{\bibfnamefont{S.~Y.} \bibnamefont{Zhou}},
  \bibinfo{author}{\bibfnamefont{X.}~\bibnamefont{Qiu}},
  \bibinfo{author}{\bibfnamefont{L.}~\bibnamefont{Ding}},
  \bibinfo{author}{\bibfnamefont{C.}~\bibnamefont{Zhang}},
  \bibinfo{author}{\bibfnamefont{U.}~\bibnamefont{Patel}},
  \bibinfo{author}{\bibfnamefont{Z.~L.} \bibnamefont{Xiao}}, \bibnamefont{and}
  \bibinfo{author}{\bibfnamefont{S.~Y.} \bibnamefont{Li}},
  \bibinfo{journal}{Phys. Rev. B} \textbf{\bibinfo{volume}{80}},
  \bibinfo{pages}{024518} (\bibinfo{year}{2009}).

\bibitem[{\citenamefont{Subedi et~al.}(2008)\citenamefont{Subedi, Zhang, Singh,
  and Du}}]{Subedi}
\bibinfo{author}{\bibfnamefont{A.}~\bibnamefont{Subedi}},
  \bibinfo{author}{\bibfnamefont{L.}~\bibnamefont{Zhang}},
  \bibinfo{author}{\bibfnamefont{D.~J.} \bibnamefont{Singh}}, \bibnamefont{and}
  \bibinfo{author}{\bibfnamefont{M.~H.} \bibnamefont{Du}},
  \bibinfo{journal}{Phys. Rev. B} \textbf{\bibinfo{volume}{78}},
  \bibinfo{pages}{134514} (\bibinfo{year}{2008}).

\bibitem[{\citenamefont{Sonier et~al.}(1999)\citenamefont{Sonier, Brewer,
  Kiefl, Morris, Miller, Bonn, Chakhalian, Heffner, Hardy, and
  Liang}}]{Sonier2}
\bibinfo{author}{\bibfnamefont{J.~E.} \bibnamefont{Sonier}},
  \bibinfo{author}{\bibfnamefont{J.~H.} \bibnamefont{Brewer}},
  \bibinfo{author}{\bibfnamefont{R.~F.} \bibnamefont{Kiefl}},
  \bibinfo{author}{\bibfnamefont{G.~D.} \bibnamefont{Morris}},
  \bibinfo{author}{\bibfnamefont{R.~I.} \bibnamefont{Miller}},
  \bibinfo{author}{\bibfnamefont{D.~A.} \bibnamefont{Bonn}},
  \bibinfo{author}{\bibfnamefont{J.}~\bibnamefont{Chakhalian}},
  \bibinfo{author}{\bibfnamefont{R.~H.} \bibnamefont{Heffner}},
  \bibinfo{author}{\bibfnamefont{W.~N.} \bibnamefont{Hardy}}, \bibnamefont{and}
  \bibinfo{author}{\bibfnamefont{R.}~\bibnamefont{Liang}},
  \bibinfo{journal}{Phys. Rev. Lett.} \textbf{\bibinfo{volume}{83}},
  \bibinfo{pages}{4156} (\bibinfo{year}{1999}), \bibinfo{note}{and references
  therein}.

\end{thebibliography}

\end{document}